# Simultaneous Manipulation of Electric and Thermal Fields via Combination of Passive and Active Schemes

Chuwen Lan[1,2], Bo Li[2]*, Ji Zhou[1]*

[1]*State Key Laboratory of New Ceramics and Fine Processing, School of Materials Science and Engineering, Tsinghua University, Beijing 100084, China*

[2]*Advanced Materials Institute, Shenzhen Graduate School, Tsinghua University, Shenzhen, China*

*Corresponding author: zhouji@mail.tsinghua.edu.cn, libo@mail.tsinghua.edu.cn,

**Abstract:**

Increasing attention has been focused on the invisibility cloak due to its novel concept for manipulation of physical field. However, it is usually realized by single scheme (namely passive or active scheme) and limited to a single field. Here, we proposed a new method to achieve simultaneous manipulation of multi-physics field via combination of passive and active schemes. Experimentally, this method was demonstrated by manipulating electric field and thermal field simultaneously. Firstly, a device was designed to behave as electric and thermal invisibility cloak simultaneously. Secondly, another device was demonstrated to behave as electric invisibility cloak and thermal concentrator simultaneously. Our method can also be extended to the other multi-physics fields, which would create much more freedom to design of new system and might enable new potential application in broad areas.

**Main text:**

Achieving the manipulation of electromagnetic wave (EMW) in a desired way has been a long-standing dream of researchers. The past decade has witnessed a rapid growth of interest in the field of metamaterial (MM) and transformation optics (TO). This can be attributed to their unprecedented ability of tailoring the EMW properties, as well as considerable phenomena and applications like invisibility cloak [1-8], negative refractive index [9], perfect lens [10] and etc. [11-12]. In particular, the invisibility cloak has gained tremendous attention due to its magic physic and potential military application. Since the first invisibility cloak reported in 2006 [2], considerable effort has been devoted to the realization of practical cloak in various ranges and much progress has been made [3-8]. In general, the cloak can be obtained by many approaches, which are usually classified into two schemes, namely passive and active schemes [13]. Up to now, however, almost all the reported investigations are limited to single scheme.

On the other hand, the great achievement in EMW has rapidly been extended to physical fields ranging from waves (acoustic field [14], elastic wave [15], matter wave [16] and etc.) to other fields like static magnetic field [17,18], *dc* electric field [19,20], thermal field [21-26], mass diffusion [27-29] and electrostatic field [30]. However, most of them are limited to manipulation of single physical field. Recently, a device was designed to simultaneously cloak the thermal and electric fields [31]. Using TO method and bifunctional metamaterial, a shell was theoretically designed to behave as a thermal concentrator and electrical invisibility cloak [32]. Our group has proposed a bilayer structure to obtain simultaneous and independent manipulation of electric and thermal fields by directly solving the physical field equations [33]. Since it is not based on TO, it can be easily achieved with naturally occurring materials. However, the above methods are strongly dependent on the background medium, thereby specific materials are required.

Here, we proposed a general method to achieve simultaneous manipulation of multi-physics field via combination of passive and active schemes. We further designed, fabricated and tested two devices simultaneously behaving as electric/thermal cloak and electric cloak/ thermal concentrator respectively to confirm the feasibility of proposed methodology. Our contribution is twofold: Firstly, the combination of passive and active schemes is proposed, for the first time to our knowledge, to achieve the manipulation of multi-physics field. This method can be extended to other multi-physics field system, such as EMW/ acoustic. Secondly, the two experimentally demonstrated devices possess simple configurations and only naturally occurring materials are required, thus greatly simplifying the fabrication. In addition, simultaneous manipulation of electric and thermal fields could create much more freedom to design new system which would enable new potential application.

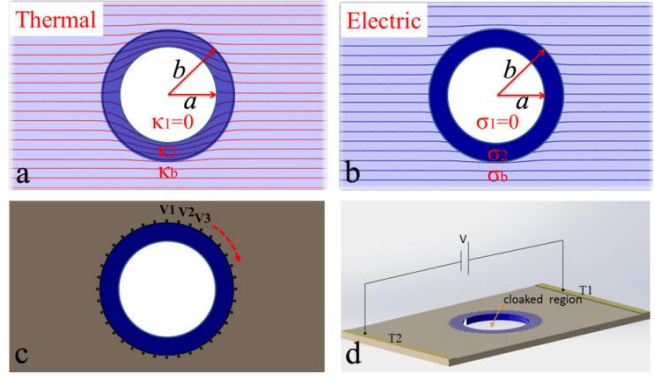

**Figure 1.** The principle for designed device behaving as thermal cloak and *dc* electric cloak: a) The thermal flux distribution. b) The current distribution. c) The schematic illustration for practical realization. d) The schematic illustration for the bifunctional cloak.

We start with a device which can simultaneously cloak the thermal and electric fields as shown in Figure 1. As depicted in Figure 1a and Figure 1b, the heat flows from the high temperature $T_1$ to the low temperature $T_2$, while the electric current is generated under a potential U=1V. An air hole with radius of *a* is placed in the center. To make this hole invisible thermally, we wrap it with a shell structure with inner and outer radii of *a* and *b* respectively. Here, the thermal conductivity and electric conductivity of the background medium are $\kappa_b$ and $\sigma_b$, while the ones for the shell are $\kappa_2$ and $\sigma_2$. It should be noted that the air hole is chosen since it can be seen as thermal and electric insulation ($\kappa_1=0$, $\sigma_1=0$). That is to say, it can protect the object placed in it from the thermal flux and electric current. According to the previous work [24,26], one can directly solve the thermal equation ($\nabla \kappa(\nabla T) = 0$) to obtain a thermal cloak, which should satisfy the follow relationship:

$$\frac{b^2}{a^2} = \frac{\kappa_2 + \kappa_b}{\kappa_2 - \kappa_b} \tag{1}$$

Thus the only task is to cloak the electric field. Similarly, the electric conductivity should satisfy the follow relationship:

$$\frac{b^2}{a^2} = \frac{\sigma_2 + \sigma_b}{\sigma_2 - \sigma_b} \tag{2}$$

However, as for the general case, $\sigma_2/\sigma_b \neq \kappa_2/\kappa_b$, then $\dfrac{b^2}{a^2} \neq \dfrac{\sigma_2+\sigma_b}{\sigma_2-\sigma_b}$. It is worth mentioning that the previous work has reported a device to cloak the electric and thermal fields simultaneously by employing specific materials and structure to make $\sigma_2/\sigma_b = \kappa_2/\kappa_b$[31]. However, it would suffer from complicate fabrication and can only work for specific case. Instead, we propose a general method to overcome those problems. As shown in Figure 2c, we applied the arrays of controlled sources around the boundary of the shell to cancel the distortion of electric field.

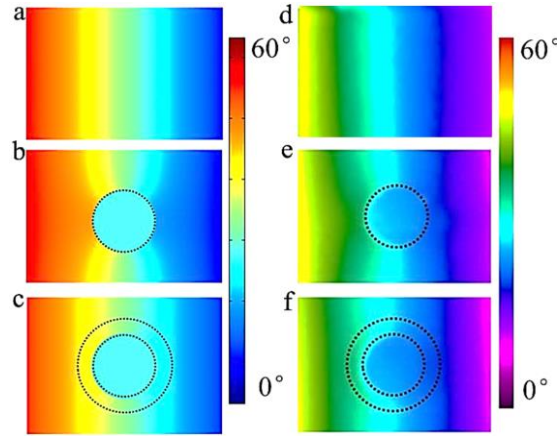

**Figure 2.** (a) Simulated temperature profile for the homogeneous medium. (b) Simulated temperature profile for the one with object (air hole). (c) Simulated temperature profile for the object (air hole) wrapped with cloak. (d) Measured temperature profile for the homogeneous medium. (e) Measured temperature profile for the one with object (air hole). (f) Measured temperature profile for the object (air hole) wrapped with cloak.

To confirm the prediction, we have designed the cloak, which is shown in Figure 2d. The inner layer is made of air ($\kappa_1$=0.02 W/mK, $\sigma_1$=0 S/m) and the outer layer is made of silicon ($\kappa_1$=150 W/mK, $\sigma_1$=0.1 S/m). The germanium (Ge) ($\kappa_1$=65 W/mK, $\sigma_1$=1 S/m) is chosen as background material. Consider $a$=5.00mm, then one can obtain $b$=8.00mm to make a thermal cloak. To cloak the electric field simultaneously, 36 discretized controlled sources are applied around the boundary of bilayer structure. The electrodes for the controlled sources are made of 200nm thickness aluminum film wafer with radii of 0.2mm. The electrodes are prepared by traditional DC magnetron magnetron sputtering system. Simulations based on Comsol Multiphysics are carried out to character the thermal conduction and electric conduction. The two sides of the sample are connected to hot source (hot water with temperature 60 ℃) and cold source (water-ice mixture with temperature 0 ℃). As for comparison, we also simulate the cases for the homogeneous background medium and the one with an air hole, as shown in Figure 2a-b. It can be found that a uniform temperature gradient is generated, while the presence of an air hole cases serious distortion of temperature. However, when the air hole is wrapped with a layer of cloak, the distortion can only occur in the cloak while the outside temperature gradient is recovered as if nothing happens (Figure 2c). Experimentally, the infrared heat camera (Fluke Ti300) was used to measure the temperature profile distribution. The measured results are shown in Figure 2d-f, showing good agreement with the simulated ones, confirming the good thermal cloak performance.

As for electric simulation, we firstly simulated the homogeneous background medium potential profile distribution, as shown in Figure 3a. Meanwhile, we also obtained the potential distribution around the bilayer boundary $V_a$. The simulated result for the bilayer structure is provided in Figure 3b, where the potential is seriously distorted. We employ the discretized sources to make the potential in corresponding positions return to be $V_a$ (The detailed voltage for each source is shown in Table 1). The simulated result is shown in Figure 3c, where the distortion of the potential is greatly decreased and nearly perfect cloak performance can be obtained. It should be noted that the imperfect cloak can be contributed to the discretized sources. The performance can be optimized by increasing the numbers of sources.

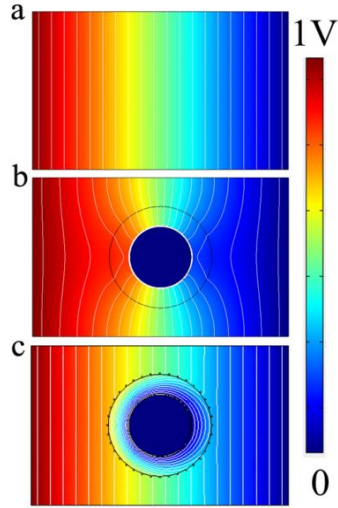

Figure 3. (a) Simulated electric potential distribution for the homogeneous medium. (b) Simulated electric potential distribution for the bilayer structure. (c) Simulated electric potential distribution for the bilayer structure with controllable sources. The isopotential lines are marked in white.

| Source | Voltage (V) | Source | Voltage (V) |
| --- | --- | --- | --- |
| 1 | 0.500 | 19 | 0.500 |
| 2 | 0.465 | 20 | 0.535 |
| 3 | 0.431 | 21 | 0.569 |
| 4 | 0.399 | 22 | 0.601 |
| 5 | 0.370 | 23 | 0.630 |
| 6 | 0.345 | 24 | 0.655 |
| 7 | 0.324 | 25 | 0.676 |
| 8 | 0.310 | 26 | 0.691 |
| 9 | 0.305 | 27 | 0.700 |
| 10 | 0.297 | 28 | 0.701 |
| 11 | 0.305 | 29 | 0.700 |
| 12 | 0.310 | 30 | 0.691 |
| 13 | 0.324 | 31 | 0.676 |
| 14 | 0.345 | 32 | 0.655 |
| 15 | 0.370 | 33 | 0.630 |
| 16 | 0.399 | 34 | 0.601 |
| 17 | 0.431 | 35 | 0.569 |
| 18 | 0.465 | 36 | 0.535 |

Table 1 The required voltage at each source for the cloaking

The performance of electric cloak can be evaluated by the potential distribution along the lines $x=-9$mm and $x=9$mm near the bifunctional cloak. In the measurement, a Multimeter (Agilent 34410A, 6, 1/2Digit Multimeter) was employed to obtain the potential. Clearly, the isopotential lines in homogeneous material are straight and parallel to each other (as shown in Figure 3a). The simulation and measured results of potential distribution along the lines ($x$=-9mm and $x$=9mm) are shown in Figure 4, where (a) and (b) correspond to line $x$=-9mm (forward scattering) and $x$=9mm (backward scattering) respectively. As expected, the employment of bilayer structure cannot make the object invisible since $\frac{b^2}{a^2} \neq \frac{\sigma_2 + \sigma_b}{\sigma_2 - \sigma_b}$. Thus, the straight isopotential lines are seriously distorted. When the bilayer structure is wrapped with array controllable sources, the isopotential lines along the lines restore to be straight ones. The measurements show good agreement with the simulation, thus confirming the feasibility of our proposed method.

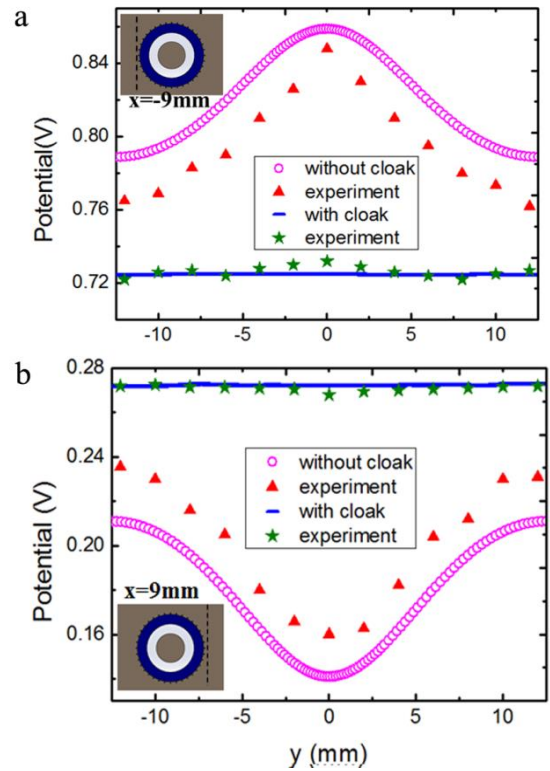

Figure 4. Simulation and experimental results of bifunctional cloak. (a) Electric potential distribution at the line x=-9mm (backward scattering). (b) Electric potential distribution at the line x=9mm (forward scattering).

To further demonstrate the robustness of our proposed method, we design a device that can behave as thermal concentrator and electric cloak simultaneously. To concentrate the thermal flux into the core region, material with homogeneous but anisotropic thermal conductivity is required. One can employ multi-layer structure to achieve this goal. The practical realization of such concentrator is shown in Figure 5, where the fan-shape layer is employed. The germanium (Ge) is chosen as background medium. 15 aluminum nitride wedges (AlN) (with thermal conductivity $\kappa=270$ W/m·K and electric conductivity $\sigma=0$ S/m) and 15 air wedges are used to construct the fan-shape layer. Due to the insulation of the fan-shape layer, the core region can be protected from the external current. Consequently, the only task is to cancel the distortion of external current. Similarly, arrays of controllable sources are used.

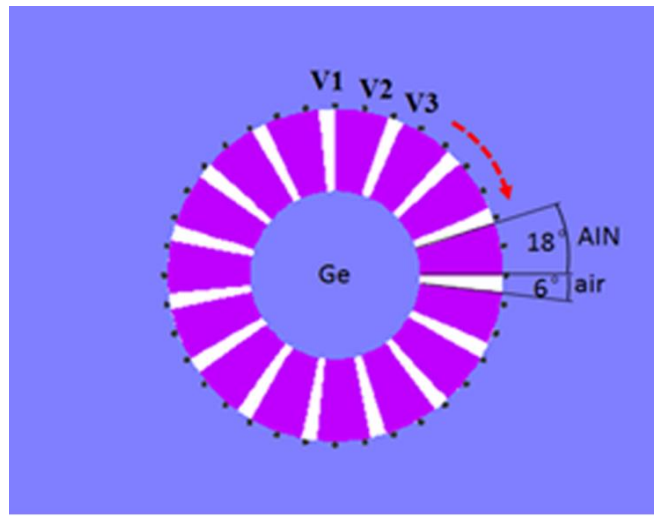

Figure 5. The schematic illustration for practical realization of designed device behaving as thermal concentrator and *dc* electric cloak.

Simulations were carried out to character the corresponding thermal and electric properties. As expected, a uniform temperature gradient (1.5 ℃/mm) is generated from the high temperature to the low temperature. The simulation for our device is shown in Figure 6a, where the thermal flux is guided into the core region resulting in the enhancement of temperature gradient (2 ℃/mm) while keeping the one for external region nearly undisturbed as nothing happens. Thus, this device behaves as a thermal concentrator. The measured result is shown in Figure 6b, which shows good agreement with the simulated one. It should be noted that the performance of the concentrator can be enhanced by increasing the anisotropy of the thermal conductivity of the fan-shape structure. To cloak the electric current field, we applied the arrays of controlled sources around the boundary of the fan-shape structure to cancel the distortion of electric field. The simulated result for the homogeneous material is shown in Figure 7a, while the one for the fan-shape structure is provided in Figure 7b where the potential is seriously distorted. According to the simulated potential profile distribution in the homogeneous background medium, as shown in Figure 7a, we can obtain the potential distribution around the bilayer boundary $V_a$. The discretized sources are employed to make the potential in corresponding positions return to be $V_a$. (The detailed voltage for each source is the same to Table 1). The simulated result is shown in Figure 7c, where good cloak performance is achieved.

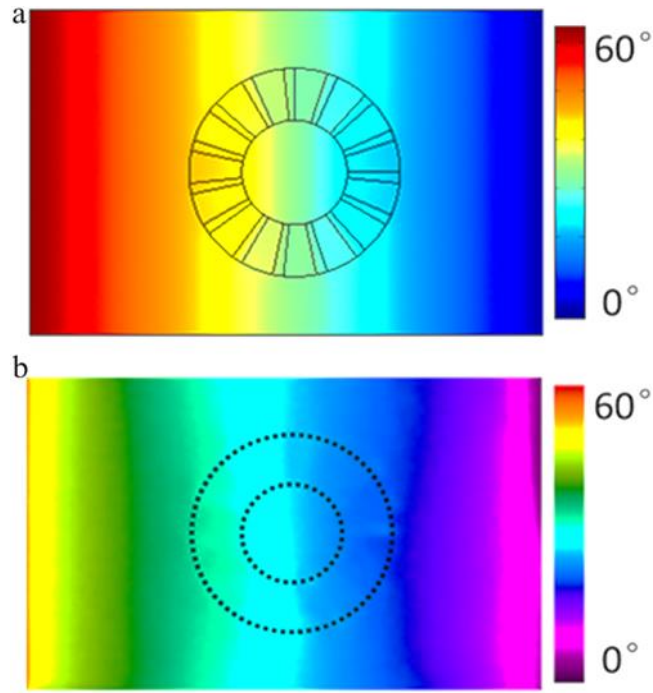

Figure 6. (a) Simulated temperature profile. (b) Measured temperature profile.

Similarly, the performance of electric cloak was evaluated by the potential distribution along the lines $x=-9mm$ and $x=9mm$ near the device (Figure 8). As expected, the original straight isopotential lines (both for the forward scattering backward scattering) are seriously distorted for the fan-shape structure without active cloak. However, when it is wrapped with arrays of controllable sources, the distortion can be cancelled remarkably and becomes straight again. It is found that the simulation and measured results show good agreement, thus confirming the robustness of our proposed method again.

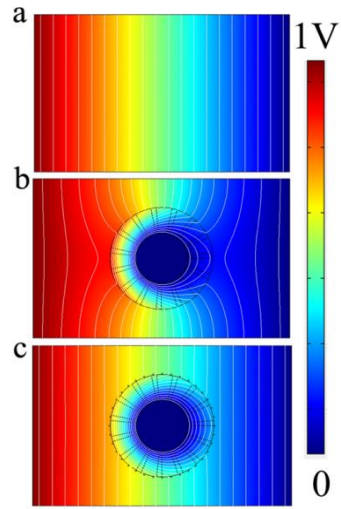

Figure 7. (a) Simulated electric potential distribution for the homogeneous medium. (b) Simulated electric potential distribution for the fan-shape structure. (c) Simulated electric potential distribution for the fan-shape structure with controllable sources. The isopotential lines are marked in white.

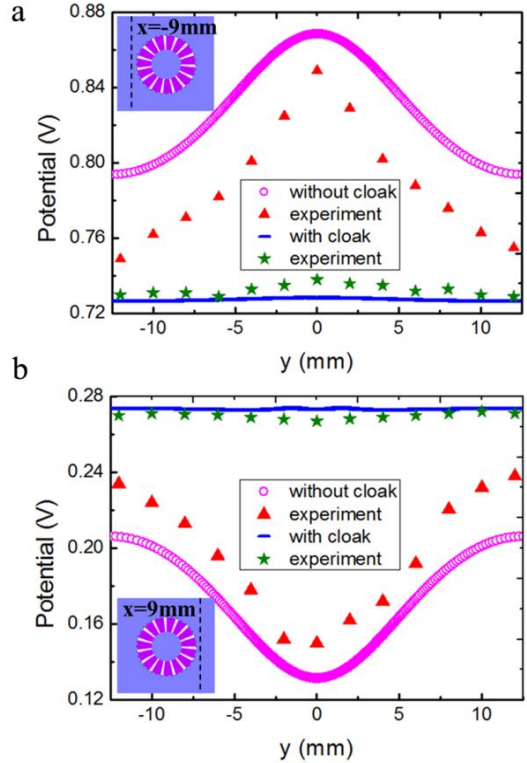

Figure 8. Simulation and experimental results of bifunctional device. (a) Electric potential distribution at the line x=-9mm (backward scattering). (b) Electric potential distribution at the line x=9mm (forward scattering).

In summary, we have demonstrated the experimental realization of simultaneous manipulation of thermal and electric current fields. The combination of passive and active schemes was proposed to achieve this goal and was demonstrated to possess many advantages like simple configuration, dynamic tunability and etc. In addition, this concept can also be extended to other multi-physics systems, which would create much more freedom to design of new system and might enable new potential application in broad areas.

This work was supported by the National Natural Science Foundation of China under Grant Nos. 51032003, 11274198, 51221291 and 61275176, National High Technology Research and Development Program of China under Grant No. 2012AA030403, Beijing Municipal Natural Science Program under Grant No. Z141100004214001, and the Science and technology plan of Shenzhen city under grant Nos.JCYJ20120619152711509, JC201105180802A and CXZZ20130322164541915.